# Preparation and Characterization of High-Entropy Alloy $(TaNb)_{1-x}(ZrHfTi)_x$ Superconducting Films

Xiaofu Zhang[1], Natascha Winter[1], Catherine Witteveen[1,2], Thomas Moehl[2], Yuan Xiao[3], Fabio Krogh[3], Andreas Schilling[1], Fabian O. von Rohr[1,2,†]

[1]Department of Physics, University of Zurich, CH-8057 Zürich, Switzerland

[2]Department of Chemistry, University of Zurich, CH-8057 Zürich, Switzerland

[3]Laboratory for Nanometallurgy, Department of Materials, ETH Zurich, CH-8093 Zürich, Switzerland

**Abstract.** We report on the preparation and the physical properties of superconducting $(TaNb)_{1-x}(ZrHfTi)_x$ high-entropy alloy films. The films were prepared by means of magnetron sputtering at room temperature, with $x$ ranging from 0 to 1 with an average thickness of 600 – 950 nm. All films crystallize in a pseudo body-centered cubic (BCC) structure. For samples with $x < 0.65$, the normal-state properties are metallic, while for $x \geq 0.65$ the films are weakly insulating. The transition from metallic to weakly insulating occurs right at the near-equimolar stoichiometry. We find all films, except for $x = 0$ or 1, to become superconducting at low temperatures, and we interpret their superconducting properties within the Bardeen–Cooper–Schrieffer (BCS) framework. The highest transition temperature $T_c$ = 6.9 K of the solid solution is observed for $x \sim 0.43$. The highest upper-critical field $B_{c2}(0)$ = 11.05 T is found for the near-equimolar ratio $x \sim 0.65$, where the mixing entropy is the largest. The superconducting parameters derived for all the films from transport measurements are found to be close to those that are reported for amorphous superconductors. Our results indicate that these films of high-entropy alloys are promising candidates for superconducting device fabrication.

## Introduction

High-entropy alloys (HEAs) are a new class of materials, which have pseudo crystalline lattices that are stabilized by a high configurational mixing entropy rather than a formation enthalpy [1,2]. Hence, due to their high entropy of constituent-mixing and the resulting minimized Gibbs free energy, they form simple crystallographic lattices. They commonly crystallize on pseudo body- or face-centered cubic (BCC or FCC) or hexagonal-closed packed (HCP) lattices with mixed site occupancies, despite their complex compositions [4–7]. The constituent atoms in HEAs are randomly distributed on the crystallographic sites, resulting in high chemical disorder. They are therefore often referred to as metallic-glasses on ordered lattices [3-8]. These alloys are currently the focus of significant attention in material science because of their versatile electronic and mechanical properties, which make them promising candidates for a large variety of applications [9]. Among other properties, HEAs have been shown to display a high fracture toughness at cryogenic temperatures and a very high strength while remaining remarkably ductile [10]. These mechanical properties are also of great interest for alloy superconductors, since they may lead to improved devices, especially under extreme conditions [9-13].

According to Anderson's theorem, weak disorder cannot destruct the pair correlations in conventional superconductors [14]. However, it is demonstrated that strong disorder, indeed, leads to spatial fluctuations of the order parameter, which may localize the Cooper pairs and eventually lead to the destruction of the superconducting state [15-19]. Nonetheless, a variety of HEAs have been reported to be superconductors, despite their exceptionally high, intrinsic degree of disorder. All, so far, reported HEA superconductors are considered to be conventional type-II superconductors with critical temperatures up to $T_c \sim 9.6$ K and upper-critical fields of up to $B_{c2}(0) \sim 11.7$ T [9]. Bulk superconducting HEAs have been found in the Ta-Nb-Zr-Hf-Ti based systems with pseudo BCC-type lattices [20-26], the pentanary $(RhPd)_x(ScZrNb)_{1-x}$ and hexanary $(RhPd)_x(ScZrNbTa)_{1-x}$ systems with CsCl-type lattices [27], the pentanary $(MoReRu)_x(ZrNb)_{1-x}$, $Re_x(HfTaWIr)_{1-x}$, and $Re_x(HfTaWPt)_{1-x}$ systems with BCC, HCP, and α-Mn-type lattices [28].

Of particular interest are the Ta-Nb-Zr-Hf-Ti HEA superconductors. An optimally doped bulk sample of the $(TaNb)_{1-x}(HfZrTi)_x$ series has very recently been reported to display a remarkably robust zero-resistance superconductivity under pressures up to $p$ = 190.6 GPa [29,30]. Furthermore, the superconductivity in this system can be well controlled and understood in terms of the respective electron count and the chemical complexity of the constituents [21].

The synthesis methodology of HEA superconductors, which is currently mostly based on the arc-melting technology in inert atmosphere, significantly limits the possibility for superconducting electronic device fabrication. Here, we demonstrate how high-quality, crystalline HEA films of the whole $(TaNb)_{1-x}(ZrHfTi)_x$ series can be obtained by a straightforward room-temperature magnetron sputtering approach. All prepared films of the $(TaNb)_{1-x}(HfZrTi)_x$ series, other than $x$ = 0 or 1, are found to be superconductors by detailed transport measurements. The superconducting parameters derived for all the films are found to be close to the parameters usually reported for amorphous superconductors. Our results strongly indicate that films of Ta-Nb-Zr-Hf-Ti HEA superconductors are promising candidates for the fabrication of superconducting devices.

**Experimental**

The $(\text{TaNb})_{1-x}(\text{ZrHfTi})_x$ ($x$ in at. %) films are deposited by direct current (DC) co-sputtering (PVD Products, Inc.) of TaNb and ZrHfTi targets (99.9%, MaTeck GmbH) at ambient temperature. By controlling the DC power of each target separately, films with different stoichiometries are deposited in a controlled fashion. The film thicknesses were determined by the *Dektak* surface step profiler.

Resistivity measurements were performed in a Quantum Design PPMS (9T) using a standard four-probe technique. Contacts are made by using 25-μm-diameter aluminium wire. The wires are contacted to the films by using the TPT wire bonder.

The phase purity and structural parameters were characterized by X-ray diffraction (XRD) in a Bragg-Brentano geometry, using a PANalytical XPert3 MRD and a Rigaku SmartLab with Cu $K_\alpha$ radiation. The elemental compositions are determined by means of energy-dispersive X-ray (EDX) analysis integrated into the Zeiss Supra 50 VP scanning-electron microscope (SEM). The compositions $x$ in this publication correspond to the values obtained from these EDX analyses. The atomic-force microscopy (AFM) was performed on an Asylum Research AFM (MFP-3D). The surface morphology was measured in tapping mode. The probe used for the measurement was a HQ:NSC15/Al BS from MikroMasch.

**Results and Discussion**

**Structural characterization of the HEA films.**

High-temperature synthesis methods, such as arc-melting, are commonly used for the preparation of HEA superconductors. These methods are prone to form intermetallic phases instead of randomly mixed lattice sites of HEA [22]. Hence, very high melting temperatures and high quenching rates have to be employed for the preparation of single phase HEAs, which correspondingly lead to a low crystallinity of the samples. Here, we have employed ambient temperature DC co-sputtering of TaNb and ZrHfTi targets to form highly crystalline $(\text{TaNb})_{1-x}(\text{ZrHfTi})_x$ films with the stoichiometries $x$ = 0, 0.04, 0.13, 0.21, 0.33, 0.43, 0.54, 0.65, 0.76, 0.88, and 1. In order to suppress any formation of competing phases, the substrates were intentionally kept at room temperature during the sputtering deposition. The atoms are quenched as soon as they arrive on the substrate, hence the sputtered films are stochastically distributed on the crystallographic positions.

In the inset of Fig. 1(a), we show the representative X-ray diffraction (XRD) pattern of the obtained $(\text{TaNb})_{0.57}(\text{ZrHfTi})_{0.43}$ film. The XRD pattern of all films can be very well indexed using the space group $Im\bar{3}m$, demonstrating the validity of a pseudo BCC structure. The observed diffraction angles of the reflections are in very good agreement with earlier works [20,21]. The diffraction peaks are broad compared to the highly crystalline reflections of the Si substrate. This effect can be attributed to the distorted lattice or the high chemical-disorder. The reflections of the HEA films are, however, sharper than the reflections obtained from HEA superconductors by arc melting [21]. A measure for this is the comparison of the full width at half maximum (FWHM) of the reflections. Given that the arc-melted samples of the earlier study and our films have been prepared very differently for the measurements and are measured on two different devices, the exact values have to be interpreted cautiously. However, the FWHM of the films is by almost a factor of 2 smaller for all the samples. Explicitly, they the ΔFWHM of the 110 reflection is approximately 0.28°, 0.25°, and 0.26° for the samples with $x$ = 0.3, 0.45, and 0.65, respectively. With these values it becomes obvious that the crystallinity of the films must be much higher than for the conventionally prepared high-entropy alloy superconductors.

Hence, this further indicates that the here used sputtering technique leads to samples with less disorder than the HEAs produced by quenching from high temperatures, which is in good agreement with earlier findings on different HEA films [31]. In Fig. 1(b), we show the (110) reflection for all prepared HEA superconducting films. The (110) reflection is found to monotonically shift as a function of the composition $x$. This change corresponds to the change of the lattice constant $a_{\mathrm{XRD}}$. Except for the highest $x$ investigated ($x$ = 0.76 and 0.88), the experimental cell parameters $a_{\mathrm{XRD}}$ are in good agreement with the expected lattice parameters $a_{\mathrm{mix}}$ estimated based on the Vegard's law of mixtures [32]. The resulting values for the lattice parameter are given in Table I.

The here prepared films were all deposited at a 30-rpm on a rotating substrate holder, resulting in films that are highly homogenous. In Fig. 1(c)&(d), we show the surface morphology of the film $(\mathrm{TaNb})_{0.57}(\mathrm{ZrHfTi})_{0.43}$ by means of AFM and SEM analyses, respectively. No granulates or agglomerates are observed in any of the two images. The surface roughness of the AFM measurement in Fig. 1(c) showed a mean roughness (Sa) of 5.6 Å and a mean square roughness (Sq) of 7.1 Å. Both values further indicate the high homogeneity and high quality of the prepared HEA films. The film thicknesses were determined to be between 600-950 nm as it is listed in Table I.

**Superconducting properties of the HEA films**
A recurring, fundamental question about the physical properties of the HEAs is whether their properties are a compositional average of properties from the constituent elements or whether they result from the collective interactions of the randomly distributed constituents. The latter has been shown to be the case for several examples in the past [20-30]. We have characterized the temperature-dependent resistivity between $T$ = 2 K and 300 K of the unmixed, highly disordered TaNb and ZrHfTi films, shown in Fig. 2(a). These films were prepared solely from the respective targets. The resulting TaNb film shows a metallic behavior at high temperatures, while its low-temperature behavior is clearly insulating in the zero-temperature limit. This behavior is commonly known for many extremely disordered films [33]. The ZrHfTi film displays a weakly insulating behavior over the whole temperature range, while there might be an onset to superconductivity just below the temperature range of these measurements, as indicated by a weak decrease in the resistivity. Neither of the two unmixed TaNb and ZrHfTi films displays a clear transition to superconductivity above 2 K, while the solid solution $(TaNb)_{1-x}(ZrHfTi)_x$ films do. It is then apparent that the superconducting properties of these HEA superconductors are not just a compositional mixture of all of the properties of the constituent elements. Instead, the highly disordered nature of the films creates a new emergent homogeneous superconducting state.

In Fig. 2(b), we present the temperature-dependent electrical resistivity $\rho_n(T)$ for all the prepared superconducting HEA films between T = 2 K and 300 K. The resistivities are normalized to the room-temperature resistivity $\rho(300K)$ for better comparability. Films with a low (TaNb) content, corresponding to a high $x$ value, display a weakly insulating slope as a function of temperature, resulting in a negative temperature coefficient. For very high (TaNb) contents, corresponding to low $x$ values, we find $\rho_n(T)$ to decrease nearly linearly with decreasing temperature, which corresponds to a highly disordered metallic behavior, commonly referred to as bad metals [21]. From this change in slope, we expect a composition mediated insulator-to-metal transition to occur around $x \sim 0.65$. In Fig. 2(c), the zero-field transitions to the superconducting state for all films are depicted. The resistivities in the vicinity to the superconducting transitions are normalized as $\rho(T)/\rho(8K)$ for better comparability. In Fig. 2(d), we show the corresponding room-temperature resistivities $\rho(300K)$ and the residual-resistivity ratio (RRR) values (defined here as $\rho(300K)/\rho(8K)$), which nearly linearly decrease from 1.5 to 0.9 with increasing values of $x$. The resistivities for the ambient-temperature deposited films are slightly larger than those of the bulk HEA superconductors obtained from arc-melting methods. This may originate in the reduced dimensionality of the films as compared to the bulk samples. The RRR values of the metallic films are similar to those observed in bulk samples, indicating similar degrees of disorder. It should be noted that slightly above the transition temperature, the resistivities of all films show a narrow temperature independent region, resulting in a plateau with a width of approximately $\Delta T = 10$ K, which is likely to correspond to impurity scattering. The critical temperatures $T_c(0)$ are shown as a function of the valence electron count in the inset of Fig. 2(e), and are tabulated in Table I.

The lowest critical temperature $T_c(0)$ of the here investigated $(TaNb)_{1-x}(ZrHfTi)_x$ films is found on the (ZrHfTi)-rich side, which corresponds to high values of $x$, for $(TaNb)_{0.12}(ZrHfTi)_{0.88}$ with a critical temperature of $T_c(0) \sim 2.8$ K. For decreasing values of $x$, the critical temperature $T_c(0)$ increases monotonically until it reaches a maximum $T_c(0)$ of 6.8 K for $x \sim 0.4$, which corresponds to an electron count per atom of $e/a \sim 4.57$. Our maximum $T_c(0)$ is at a somewhat lower electron count and approximately 1 K lower than the values reported for $(TaNb)_{1-x}(ZrHfTi)_x$ bulk samples where $e/a \sim 4.7$ at $T_c(0) \approx 8$ K (in the resistivity), which is likely caused by a confined-size effect in our films. For values of $x$ smaller than 0.4, the critical temperatures monotonically decrease to a value of $T_c(0) \sim 5.5$ K for $(TaNb)_{0.96}(ZrHfTi)_{0.04}$.

**Parameters characterizing the superconducting state**
In Fig. 3(a), we present the field dependence of the resistivity of the sample $x$ = 0.4 in the vicinity of the superconducting transition. This figure represents the typical behavior of all the superconducting films investigated here. It should be noted that the superconducting transitions for these HEA films are remarkably sharp even in large magnetic fields, especially when compared with other binary alloys [34,35]. The transition temperatures shift to lower values with increasing field. We have employed the commonly used 50%-criterion of the normal-state resistivity (see dashed line in Fig. 3(a)) to extract the transition temperatures at different magnetic fields ($T_c(B)$). This procedure allows us to determine the temperature dependence of the upper-critical field $B_{c2}(T)$. In Fig. 3(b), we show the corresponding values for $B_{c2}(T)$ for all the deposited films. The dashed lines are linear fits near zero-field, determining the slope of $dB_{c2}(T)/dT$. It is interesting to note that these slopes monotonically increase with decreasing $x$, as it is shown in Fig. 3(c). The actual upper-critical field, however, deviates from the linear dependence for large applied magnetic fields. Using the Werthamer–Helfand–Hohenberg (WHH) approximation, which considers the electron-spin and spin-orbital scattering [36], we can estimate the $B_{c2}(0)$ in the dirty-limit as

$$B_{c2}^{\text{WHH}}(0) = -0.69 \cdot T_c(0) \cdot \left(\frac{dB_{c2}(T)}{dT}\right)_{T \to T_c(0)}.$$

The corresponding values of $B_{c2}^{\text{WHH}}(0)$ are plotted for all films in Fig. 3(c) and summarized in Table I. The maximum in $B_{c2}^{\text{WHH}}(0)$ is near $x \sim 0.6$. This stoichiometry corresponds to an almost equimolar elemental ratio of the constituents, where the mixing entropy $\Delta S_{\text{mix}} = -5R \sum_{i=1}^{5} x_i \ln x_i$ reaches its maximum $\Delta S_{\text{mix}} \sim R \ln 5$, with the mole fractions $x_i$ of each component, and the gas constant $R$ [20].

In Fig. 3(c), we also show the Pauli paramagnetic limit $B_{\text{Pauli}}$ for all stoichiometries. This limit can be estimated according to 1.84×$T_c$(0) for an isotropic s-wave spin-singlet in the weak coupling BCS case [37]. Different from the WHH model in which pair condensation is suppressed due to the Lorentz force from the external field on acting the opposite spins, the Pauli paramagnetic limit originates from a spin-pair breaking mechanism, and normally $B_{c2}^{\text{WHH}}(0)$ is lower than the Pauli paramagnetic limit. However, for superconductors with strong electron-phonon coupling or spin-orbit coupling, the upper-critical field can surpass this limit [38]. It is interesting to note that on the TaNb-poor side $x > 0.6$, $B_{c2}^{\text{WHH}}(0)$ exceeds the Pauli paramagnetic limit by up to 20%, while on the TaNb-rich side, the upper-critical fields are below this limit. This effect is much more pronounced in the here investigated thin films than in the bulk HEA superconductors [16]. The exotic high upper-critical field for some of these films may be attributed to strong electron-phonon coupling, since enhanced spin-orbit coupling effects appear to be unlikely [21].

We have derived the basic parameters characterizing the superconducting state to gain more insights into these superconducting HEA films. Based on the framework of Ginzburg-Landau (GL) theory, the zero-temperature GL coherence length $\xi_{GL}(0)$ is related to the zero-temperature upper-critical field via $\xi_{GL}(0) = [\Phi_0/2\pi B_{c2}(0)]^{0.5}$ [34]. The corresponding values for $\xi_{GL}(0)$ as obtained from the extrapolated $B_{c2}^{\text{WHH}}(0)$ are depicted in Fig. 4(a). The superconducting coherence length corresponds to the length scale over which the superconducting order parameter can be affected by a local or external perturbation. It is, therefore, plausible that the minimum of $\xi_{GL}(0)$ is realized for the composition where the maximum mixing entropy is reached, i.e., for maximum disorder, around $x \approx 0.6$.

Another fundamental parameter is the magnetic penetration depth $\lambda$, over which magnetic fields decay exponentially near the interface between a superconducting and a normal-state interface. In the dirty limit, the zero-temperature magnetic penetration depth $\lambda(0)$ is expressed as

$$\lambda(0) = [\hbar \rho_n / \pi \mu_0 \Delta(0)]^{0.5}$$

where $\hbar$ is the Planck constant, $\mu_0$ is the vacuum permeability, and $\Delta(0)$ is the zero-temperature superconducting energy gap [34]. In the BCS weak-coupling limit, this energy gap is $2\Delta \sim 3.5 k_B T_c$ [39], which has also be experimentally confirmed for Ta$_{34}$Nb$_{33}$Zr$_{14}$Hf$_8$Ti$_{11}$ [20]. We have therefore estimated the values for $\Delta(0)$ of our films by applying this BCS relationship. From the normal state resistivities in Fig. 2(c) we obtained the values for $\lambda(0)$, which are shown in Fig. 4(a). The respective values for the zero-temperature GL parameter $\kappa = \lambda/\xi$ are displayed in Fig. 4(b). We find $\kappa$ to decrease monotonically with decreasing $x$. It should be noted that all values for $\kappa$ for all the here investigated films are far larger than $1/\sqrt{2}$, thereby classifying them as strongly type-II superconductors. Another parameter that significantly influences the properties of superconducting devices is the diffusion constant of quasiparticles or normal state electrons $D_e$ [34]. The $D_e$ can be determined from the data for slope of $B_{c2}(T)$ in Fig. 3(b) according to [40],

$$D_e = \frac{4 k_B}{\pi e} \cdot \left(\frac{dB_{c2}}{dT}\right)^{-1}$$

where $k_B$ is the Boltzmann constant, and they are plotted in Fig. 4(b) as a function of $x$. It is interesting to note that the parameters characterizing the superconducting state of these HEA films are rather close to those of amorphous superconducting films, such as WSi or MoGe [34,41]. Such amorphous films have been found to be exceptionally promising for the fabrication of superconducting nanowire single-photon detectors [42-44] since they are robust against local constrictions [44,45]. Therefore, the highly disordered nature of superconducting HEA films makes them also good candidates for device fabrication.

**Summary and Conclusion**

We have successfully deposited a series of high-quality superconducting $(TaNb)_{1-x}(ZrHfTi)_x$ HEA films by magnetron sputtering at ambient-temperature. The XRD pattern for these films demonstrate that they all arrange on pseudo-BCC crystal lattices, despite the large amount of disorder. These films are highly homogeneous, and display nearly amorphous electronic behavior, similar with that of the amorphous WSi or MoGe films [34,41]. For high values of $x$, we have found a weakly insulating behavior in the normal state, while samples with higher values of $x$ are found to be bad metals.

A weak-insulator to bad-metal transition as a function of chemical composition is found to occur around $x \sim 0.65$. The films consisting only of TaNb or ZrHfTi are not superconducting above 2 K. However, all the mixed films of the $(TaNb)_{1-x}(ZrHfTi)_x$ HEA solid solution are strongly type-II superconductors with sharp superconducting transitions. The highest critical temperature $T_c \approx 6.9$ K is found for a stoichiometry of $x \sim 0.43$. The highest upper-critical field, however, is realized in samples with the highest mixing entropy with near-equimolar constituents. For $x \lesssim 0.7$, the upper critical field even exceeds the Pauli paramagnetic limit. Finally, we have demonstrated that the parameters characterizing the superconducting state of these HEA superconductors resemble those of amorphous superconductors. Therefore, HAE films are promising candidates for the fabrication of superconducting nanostructures and devices.

**Acknowledgements**

This work was supported by the Swiss National Science Foundation under Grant No. PZ00P2_174015.

TABLE I. The $(TaNb)_{1-x}(ZrHfTi)_x$ compositions $x$ from EDX, the electron count $e/a$, the lattice parameter $a_{mix}$, the zero-field transition temperature $T_c(0)$, the WHH zero-temperature upper critical field $B_{c2}^{WHH}(0)$, and the Pauli paramagnetic limit $B_{c2}^{Pauli}(0)$ of all the investigated films.

| $x$ | $d$ (nm) | $e/a$ | $a_{mix}$ (Å) | $a_{XRD}$ (Å) | $T_c(0)$ (K) | $B_{c2}^{WHH}(0)$ (T) | $B_{c2}^{Pauli}(0)$ (T) |
|---|---|---|---|---|---|---|---|
| 0.88 | 740 | 4.12 | 3.45 | 3.53 | 2.77 | 6.15 | 5.10 |
| 0.76 | 623 | 4.24 | 3.43 | 3.47 | 4.61 | 10.44 | 8.48 |
| 0.65 | 622 | 4.35 | 3.41 | 3.43 | 5.60 | 11.05 | 10.30 |
| 0.54 | 620 | 4.46 | 3.39 | 3.42 | 6.14 | 9.93 | 11.30 |
| 0.43 | 625 | 4.57 | 3.37 | 3.39 | 6.76 | 8.77 | 12.44 |
| 0.33 | 635 | 4.67 | 3.36 | 3.37 | 6.43 | 7.05 | 11.83 |
| 0.21 | 630 | 4.79 | 3.34 | 3.36 | 6.23 | 5.78 | 11.46 |
| 0.13 | 750 | 4.87 | 3.32 | 3.34 | 6.02 | 4.29 | 11.08 |
| 0.04 | 955 | 4.96 | 3.31 | 3.34 | 5.57 | 2.95 | 10.25 |

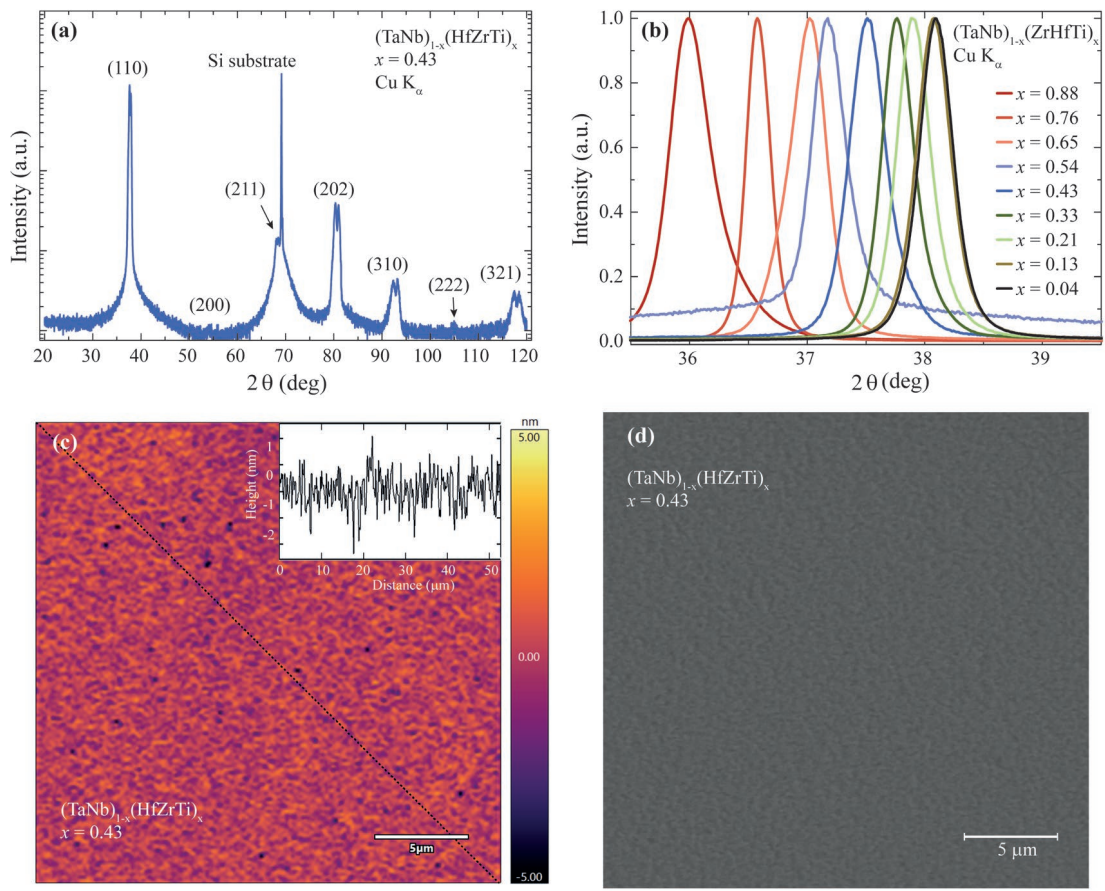

Figure. 1 (a) X-ray diffraction pattern of the $x = 0.43$ film. The peaks are indexed according to a pseudo BCC crystal lattice. (b) The corresponding (110) reflection for all the investigated films. Surface structure of the film $x = 0.43$ $(TaNb)_{0.57}(ZrHfTi)_{0.43}$ measured with (c) AFM and (d) SEM.

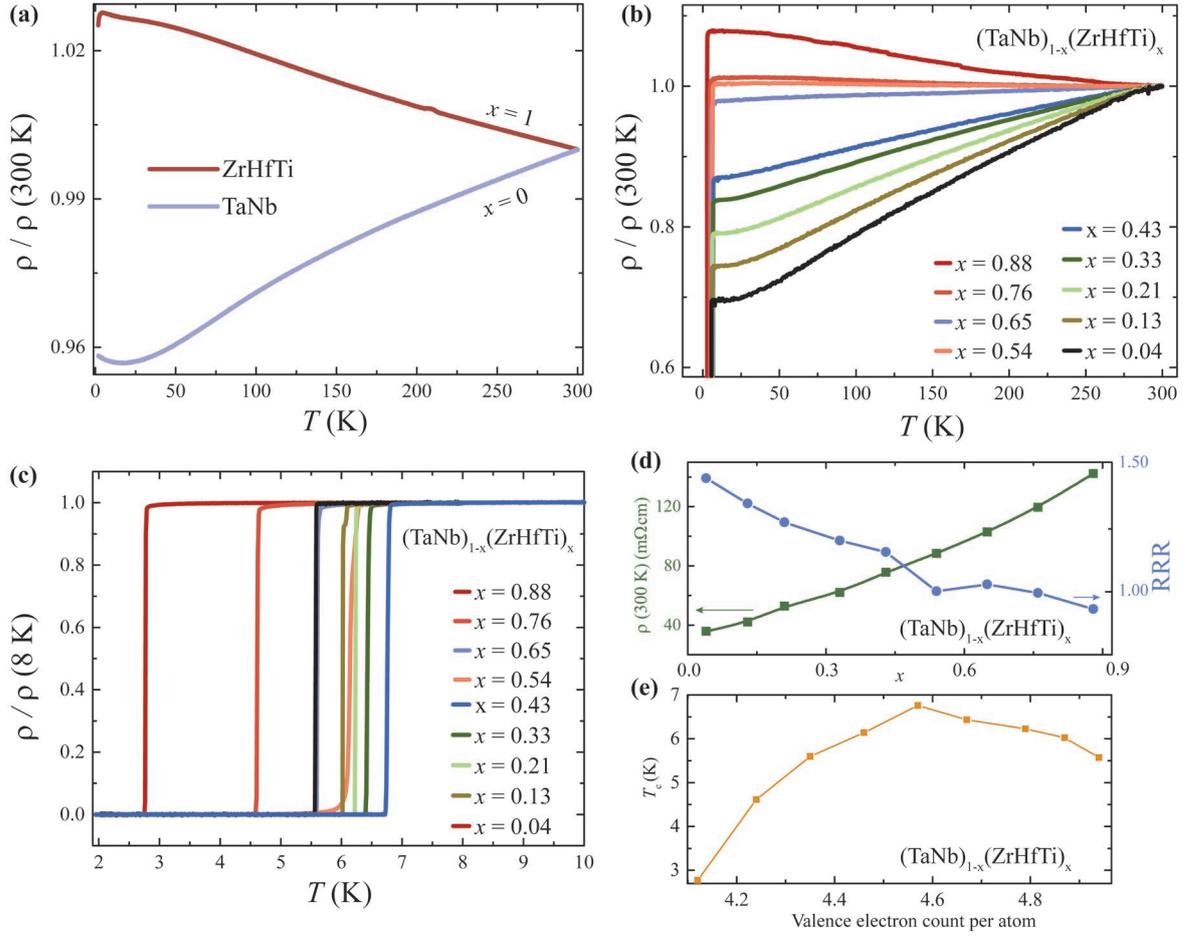

Figure 2: Temperature dependent resistivities $\rho(T)/\rho(300K)$ between 2 K and 300 K of the (a) disordered TaNb and ZrHfTi thin films and (b) of the HEA films. (c) Normalized $\rho(T)/\rho(8K)$ zero-field superconducting transitions for the HEA films between 2 K and 8 K. (d) Room-temperature resistivity and the residual resistivity ratio for all the HEA films. (e) The zero-field transition temperature as a function of the valence-electron count.

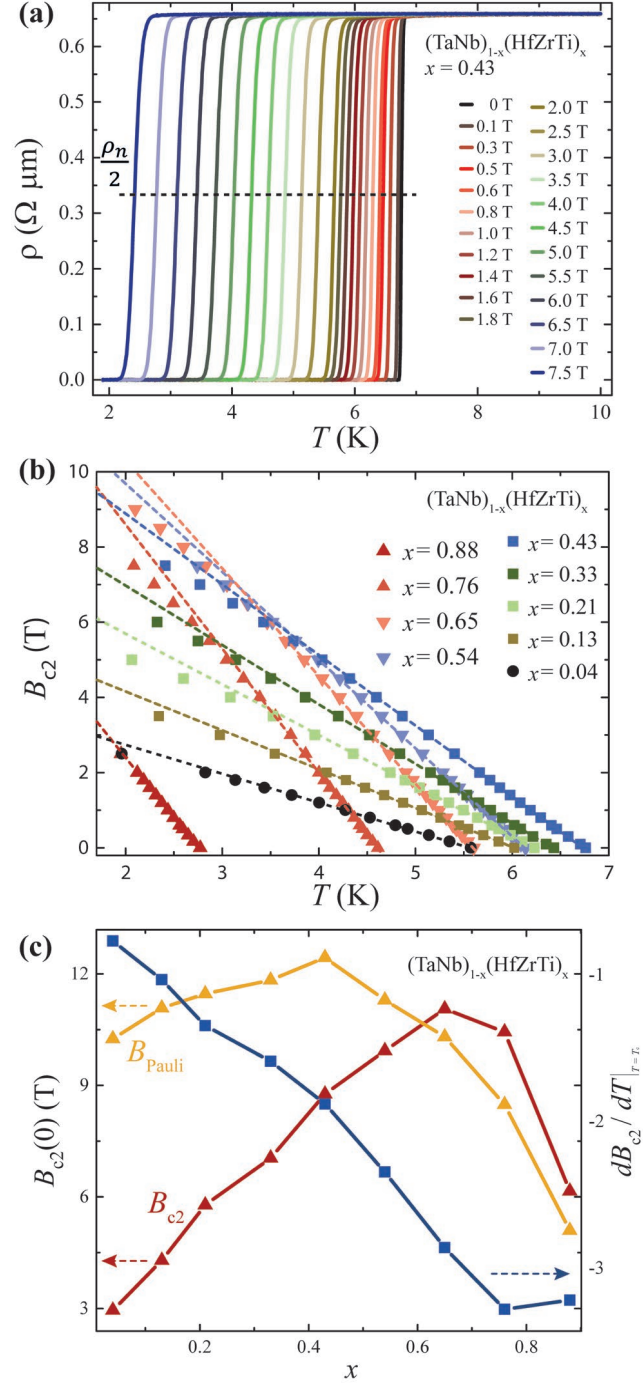

Fig. 3 (a) The magnetic field dependence of the superconducting transition in fields from 0 to 7.5 T of the $x = 0.43$ film. The dashed line illustrates the 50% of the normal-state resistivity criterion for determining the field dependent critical temperatures. (b) The temperature dependence of the upper-critical field. The dashed lines are linear fits in the zero field limit. (c) Upper-critical fields and the slopes as a function of the $x$. The dark yellow line corresponds to the Pauli paramagnetic limit.

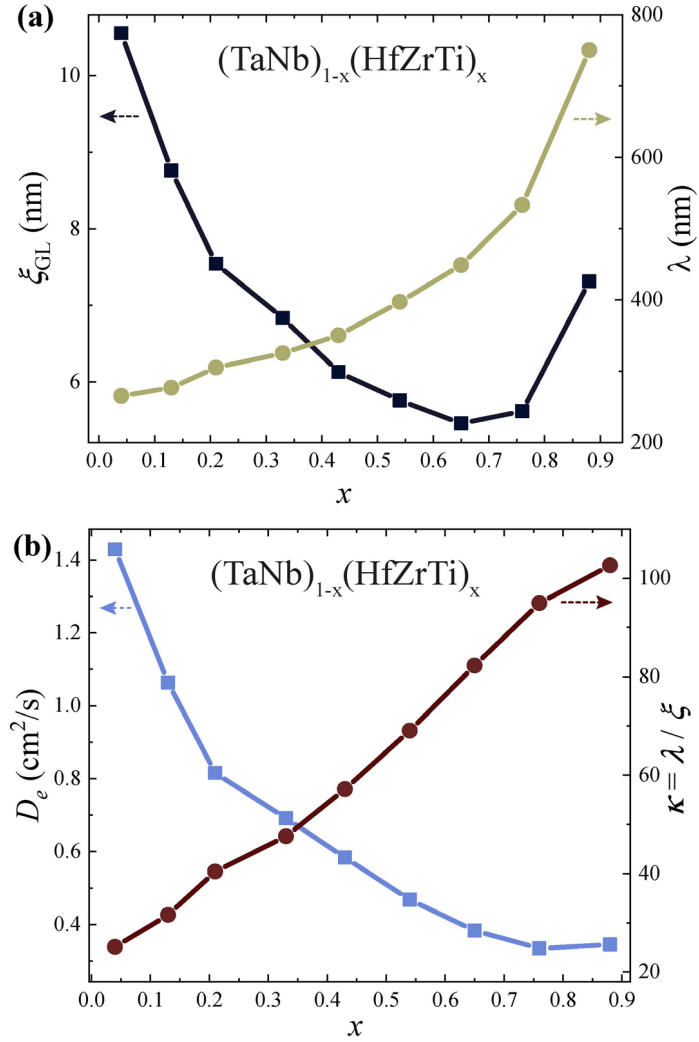

Figure 4: Parameters characterizing the superconducting state of the HEA superconducting films. (a) The zero-temperature GL coherence length $\xi_{GL}$ and the penetration depth $\lambda$ for all the films. (b) The normal state diffusion constants $D_e$ and the GL parameters $\kappa$ as functions of $x$.